\documentclass[12pt,preprint]{aastex}

\slugcomment{To appear in Astronomical Journal}

\shorttitle{Anomalous Balmer Line Strengths}
\shortauthors{Poole et al.}

\begin{document}

\title{On the Anomalous Balmer Line Strengths in Globular Clusters}

\author{Violet Poole, Guy Worthey, Hyun-chul Lee and Jedidiah Serven}
\affil{Department of Physics and Astronomy, Washington State University,
    Pullman, WA 99163}
\email{gworthey@wsu.edu}

\begin{abstract}
Spectral feature index diagrams with integrated globular clusters and simple
stellar population models often show that some clusters have weak H$\beta$, so
weak that even the oldest models cannot match the observed feature depths. In
this work, we rule out the possibility that abundance mixture effects are
responsible for the weak indices unless such changes operate to cool the
entire isochrone. We discuss this result in the context of
other explanations, including horizontal branch morphology, blue straggler
populations, and nebular or stellar emission fill-in, finding a preference for
flaring in M giants as an explanation for the H$\beta$ anomaly.
\end{abstract}


\section{Introduction}

Globular cluster are often the only objects that can be detected in the halos
of other galaxies. Since most globular cluster are very old, studying them
provides vital information about the formation and chemical history of these
galaxies \citep{gs04}. The similar ages of the constituent stars and low
velocity dispersion makes these objects relatively easy to study in
integrated light.

The most reliable way to obtain information about age and abundance patterns
of a stellar population is star by star analysis via a color-magnitude diagram
(CMD).  However this technique is only feasible for nearby objects, due to the
limits of current instrumentation. Therefore, we need to develop reliable
methods for analyzing the composite light from all the stars in these
systems. This is no easy feat, since many factors can complicate the analysis.

One of the biggest problems that plagues the study of the integrated light of
a stellar population is the very similar effects that age and metallicity have
on the spectra of stellar populations \citep{oc86}. However, the degeneracy
between the age and metallicity can be broken. Rabin (1980, 1982) and
\citet{gst81} noticed that the Balmer lines are quite sensitive to
age. Combining this with the knowledge that metal lines, such as Mg $b$,
[MgFe], or $<$Fe$>$, are relatively more sensitive to metallicity than age
\citep{w94}, the degeneracy between age and metallicity can be broken by
plotting the strength of the Balmer lines versus metallic absorption
blends. This is because H$\beta$ operates in such a way that its strength is
nonlinear with temperature, especially between 6000 K and 9000 K, where
main-sequence turnoff stars from a few hundred Myr to ancient reside. But
stars of other kinds also inhabit that temperature band.

The H$\beta$-metal-index grids give the impression that age and
metallicity are the only factors that affect a population's location
in the grid, but of course this is misleading. As you can see from Fig \ref{1}
(a), other factors must be affecting the values of the indices, since they are
quite scattered and some are off the grid.  For comparison, an average of the
Virgo elliptical galactic nuclei is also plotted along with the globular
cluster data (J. Serven et al. 2009, in preparation).

Observational difficulties play a main role, of course.  There are sometimes
horizontal branch and blue straggler stars that can contribute enough to alter
the lines strengths. Rarer stars such as AGB-Manque stars \citep{ren} and
planetary nebula can be ruled out as being significant contributers under
ordinary circumstances, as can the much fainter white dwarf population. These
warm stars make the Balmer indices stronger, not weaker, so for populations of
these stars to represent a ``solution'' to the mystery, there would also need
to be a systematic error in the models to weaken Balmer index strength.

Additionally, there is the possibility that H$\beta$ is
being filled in by nebular emission from hydrogen recombination lines.
This fill-in could come from diffuse gas, but it could also come from
flaring stars of various sorts: asymptotic giant branch (AGB) stars,
M-type dwarfs, cataclysmic variables, and others are known to have
transient emission-line spectra \citep{sc05}.

Finally, there is the possibility that abundance-mixture effects
could drive a significant change in H$\beta$ and other feature
strengths. This is a relatively unexplored avenue of investigation,
but the tools now exist to probe the question \citep{dot07, le09}, and that is
our primary task in this paper.

The models and observational data are already available in the
literature, but the implementation is recapped in $\S 2$. The
implications of our investigation are discussed in $\S 3$, and then
there is a brief conclusion.

\section{Observations and Models}

A version of integrated-light models \citep{w94,t98} that use a new grid of
synthetic spectra in the optical \citep{le09} in order to investigate the
effects of changing the detailed elemental composition on an integrated
spectrum was used to create synthetic spectra at a variety of ages and
metallicities for single-burst stellar populations. The underlying isochrones
for most of the present paper were the \citet{w94} ones, because they
allow us ``manual'' HB morphology control.  However, there are certain
caveats to using these isochrones.  Specifically the models are a bit crude by
today's standards and the ages are about 2 Gyr too old, so that 17 Gyr should
really be interpreted as 15 Gyr. Other isochrone sets were used to check the
results.

For this exercise, new stellar index fitting functions were generated. The
data sources include a variant of the original Lick collection of stellar
spectra \citep{w94b} in which the wavelength scale of each observation has
been refined via cross-correlation, as well as the MILES spectral library
\citep{miles} with some zero point corrections, and the Coude Feed library
(CFL) of \citet{cfl}. The CFL was used as the fiducial set, in the sense 
that any zero point shifts between libraries were corrected to agree with the
CFL case. The MILES and CFL spectra were smoothed to a common Gaussian
smoothing corresponding to 200 km s$^{-1}$. The rectified-Lick spectra were
measured and then a linear transformation was applied to put it on the fiducial
system. 

Multivariate polynomial fitting was done in five overlapping temperature
swaths as a function of $\theta_{eff} = 5040 / T_{eff}$, log $g$, and
[Fe/H]. The fits were combined into a lookup table for final use. As in
\citet{w94}, an index was looked up for each ``star'' in the isochrone and
decomposed into ``index'' and ``continuum'' fluxes, which added, then
re-formed into an index representing the final, integrated value after the
summation. This gives us empirical index values. After that, additive index
deltas were applied as computed from the grid of new synthetic spectra when
variations in chemical composition are needed. The grid of synthetic spectra
is complete enough to predict nearly arbitrary composition changes.

\begin{figure}
\epsscale{.80}
\plotone{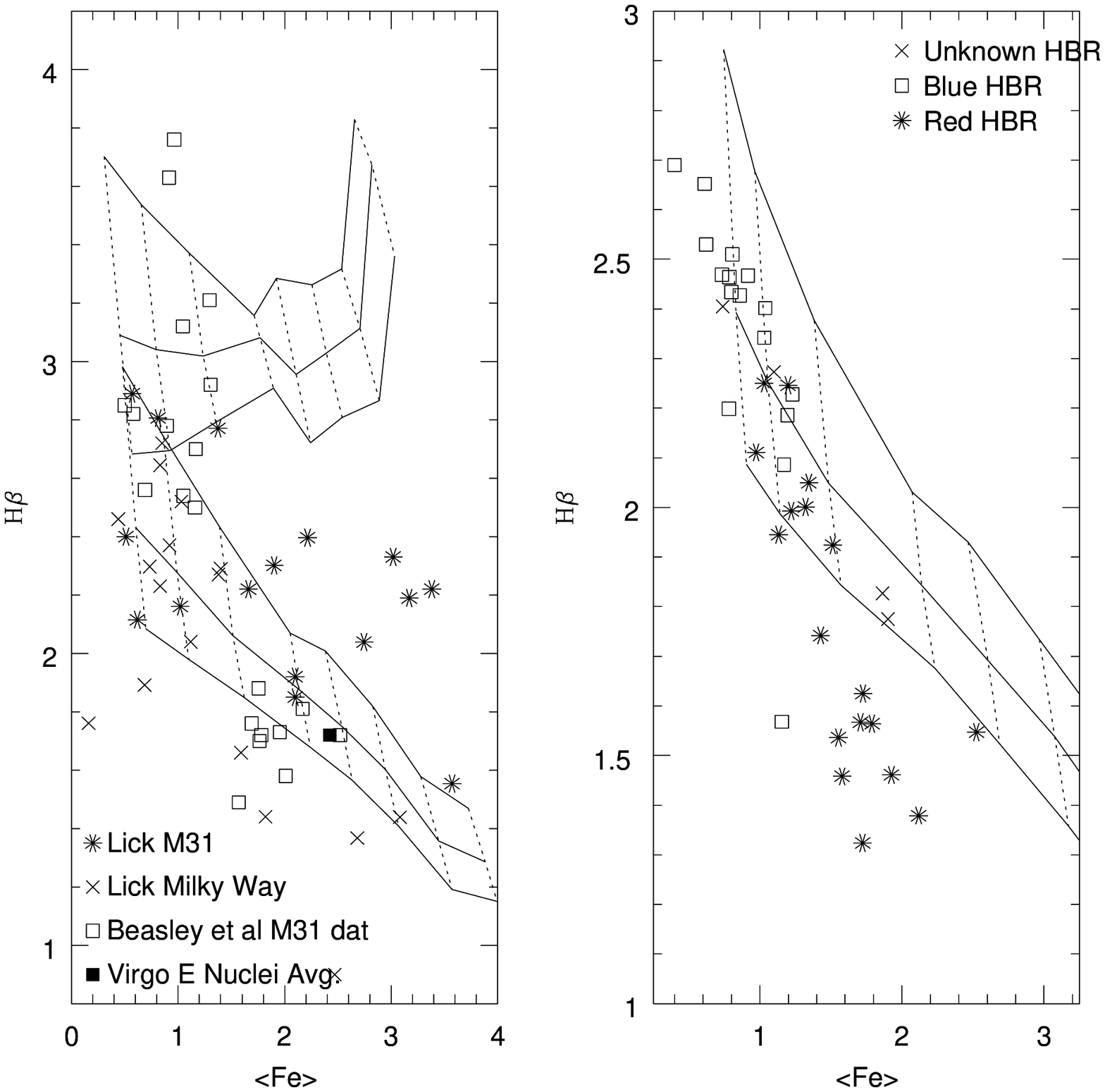}
\rotate
\caption{(a) Literature data for Lick indices plotted for globular clusters in
  the MilkyWay and M31 \citep{t98,beas}. Also plotted is an average of the
  Virgo elliptical galactic nuclei (J. Serven et al. 2009, in preparation).
  Two model grids \citep{wo97} are shown in this figure.  The lower one has a
  forced red clump morphology for the horizontal branch, while the top grid
  has all hot horizontal branch stars. Even using different models, we still
  have the problem that some globular clusters are weaker in H$\beta$ than the
  reddest models, appearing much older than could be realistic.
 (b) H$\beta$ and $<$Fe$>$ indices for globular clusters (from Schiavon et
  al. 2005) and models.  Red lines are models with ages:  8, 12, and 17 Gyrs
  from top to bottom. Blue lines are models at different
  metalicities. Globular Clusters are plotted on the same graph, divided into
  two groups.  Red HBR, have $X_{HB}=(B-R)/(B+V+R)$ greater than zero, while
  Blue HBR type have $X_{HB}$ less than zero. Both of the pair of Fe-strong
  blue X's (NGC 6388, NGC 6441) are known to have a partially blue horizontal
  branch \citep{bus}. The uncertainy is smaller than the point size used to plot
  the data. \label{1}} 
\end{figure}

\begin{figure}
\epsscale{.80}
\plotone{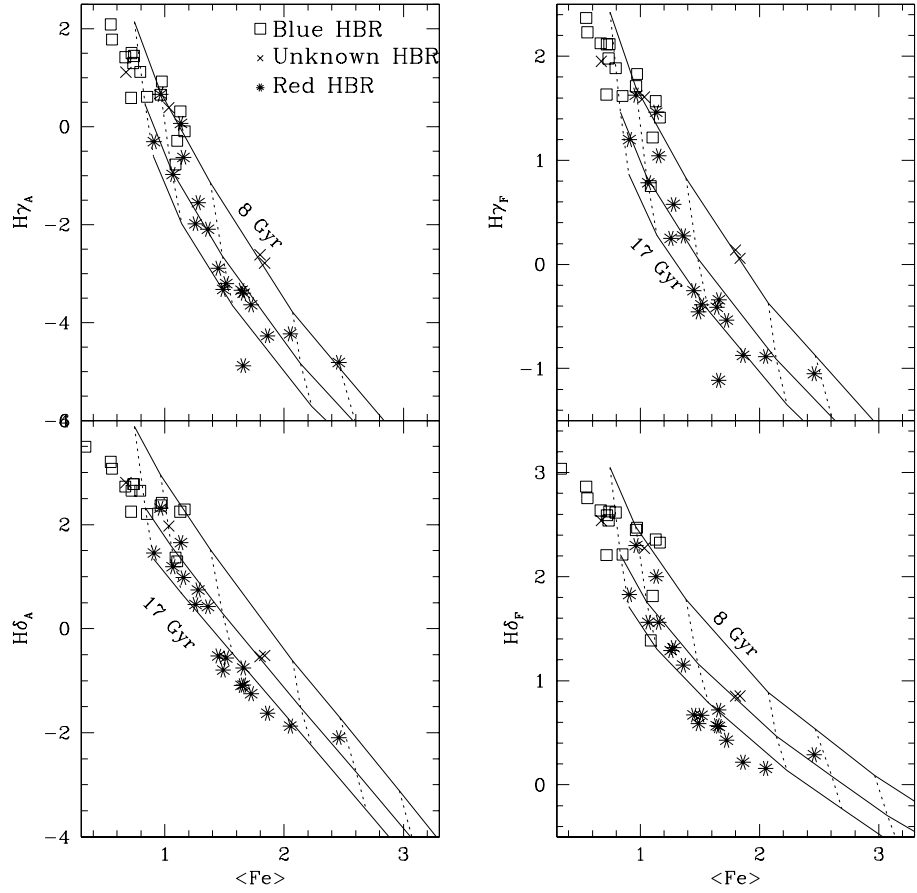}
\caption{(a)H$\gamma_A$ versus $<$Fe$>$ for globular clusters and models.
(b)H$\gamma_F$ versus $<$Fe$>$ for globular clusters and models. 
(c)H$\delta_A$ verse $<$Fe$>$ for globular clusters and models.
(d)H$\delta_F$ versus $<$Fe$>$ for globular clusters and models. The
  meanings of lines and symbols are the same as in Fig 1 (a).\label{2}} 
\end{figure}

We also smoothed the \citet{sc05} globular cluster spectra to 200 km
s$^{-1}$ or the Lick \citep{wo97} resolution, as needed, and measured
Lick or Lick-like pseudo-equivalent width indices
\citep{w94b,wo97,s05} from them. When the globular clusters and the
age-metallicity model grid of values were plotted on the same graph (see
Figures \ref{1} (b) and \ref{2} ) a globular cluster's position in the grid
allows one to estimate its age and metallicity, at least naively.

Cursory examination of these graphs yields a puzzling thing. On graphs with
H$\beta$ as one of the indices, some of the globular clusters lie much below
the oldest age plotted of 17 Gyr (see Figure \ref{1} (a) ).  However, graphs
that are not plotted with the H$\beta$ index as one of the axes do not have
this problem (see Figure \ref{2} ). This indicates that there could be
something going on in the spectra of these globular clusters near the H$\beta$
line that does not affect H$\gamma$ or H$\delta$ to the same degree. It could
also indicate that the models for the H$\beta$ index are not correct.

\section{Discussion: Balmer Features in the Integrated Light of Globular Clusters}

Many factors could potentially affect the Balmer features in the integrated
light of the globular clusters. We consider effects due to abundance ratios,
horizontal branch morphology, the presence of blue stragglers, and emission
fill-in of the Balmer lines due to hydrogen recombination lines from either
external nebulae or stellar activity in individual cluster stars. We also
consider the illusions due to miscalibrated models.

\subsection{Abundance Ratio Effects}

\begin{figure}
\epsscale{.80}
\plotone{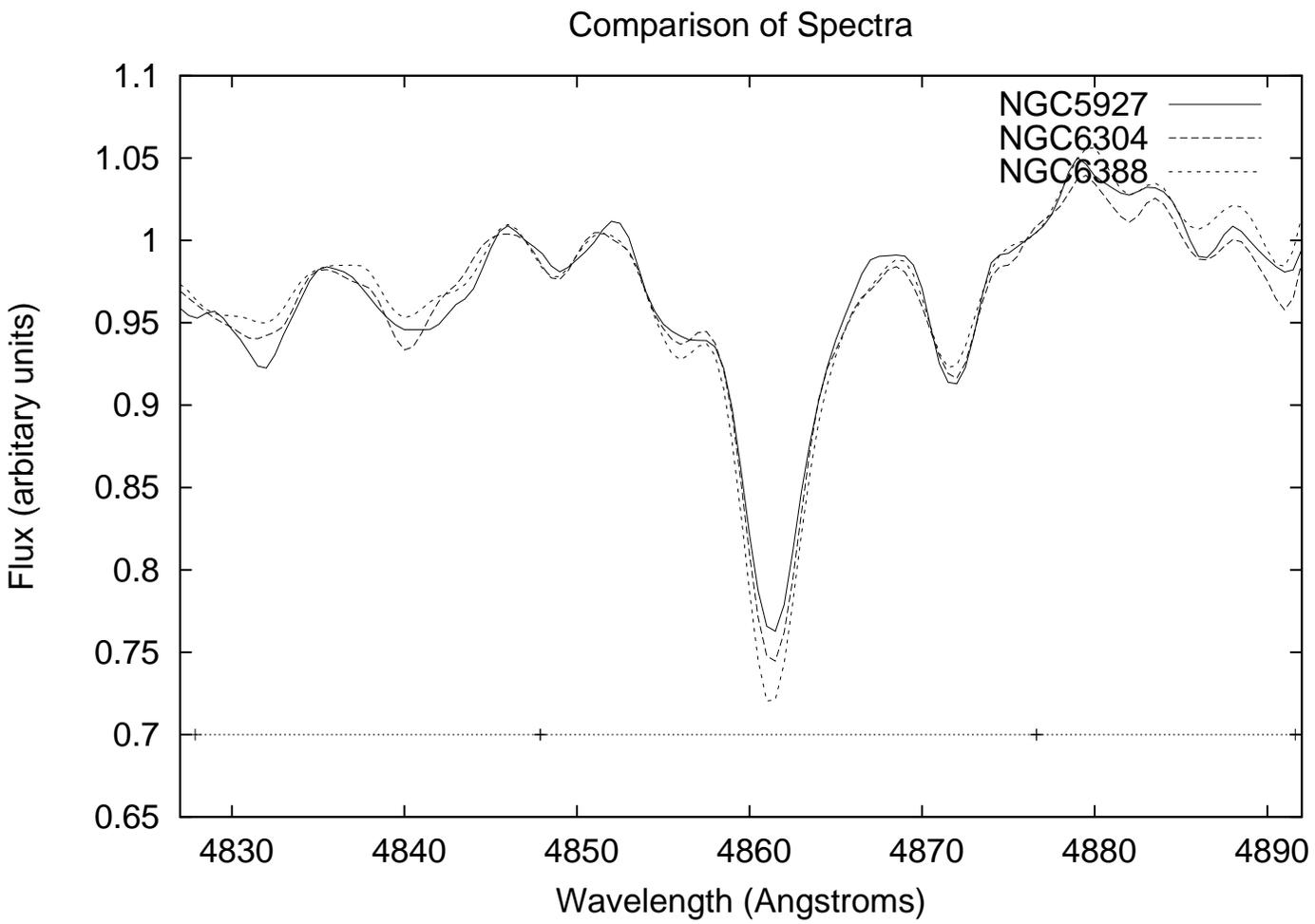}
\caption{Comparison of the spectra of globular clusters at approximately the
  same metallicity. NGC 6388 appears on the age-metal grid while NGC 5927 and
  NGC 6304 are  below the grid, and the stronger H$\beta$ feature is
  obvious. \label{3}} 
\end{figure}

It is of interest to examine the spectra themselves for evidence to
support the various hypotheses that could explain their behavior.
Comparison of the spectra of the globular clusters with ages off the
grid to the spectra of globular clusters with similar metallicity
lying within the grid indicates that the main difference is the depth
of the H$\beta$ line itself, rather than a difference in heights of
either the blue or red continuum bands (see Figure \ref{3}).  Specifically,
the H$\beta$ line of the clusters off the grid are shallower than
those that are on the grid, and, morphologically, this does not seem
to be a problem in the continuum regions at all, but a true modulation
of the H$\beta$ line itself.

\begin{figure}
\epsscale{.80}
\plotone{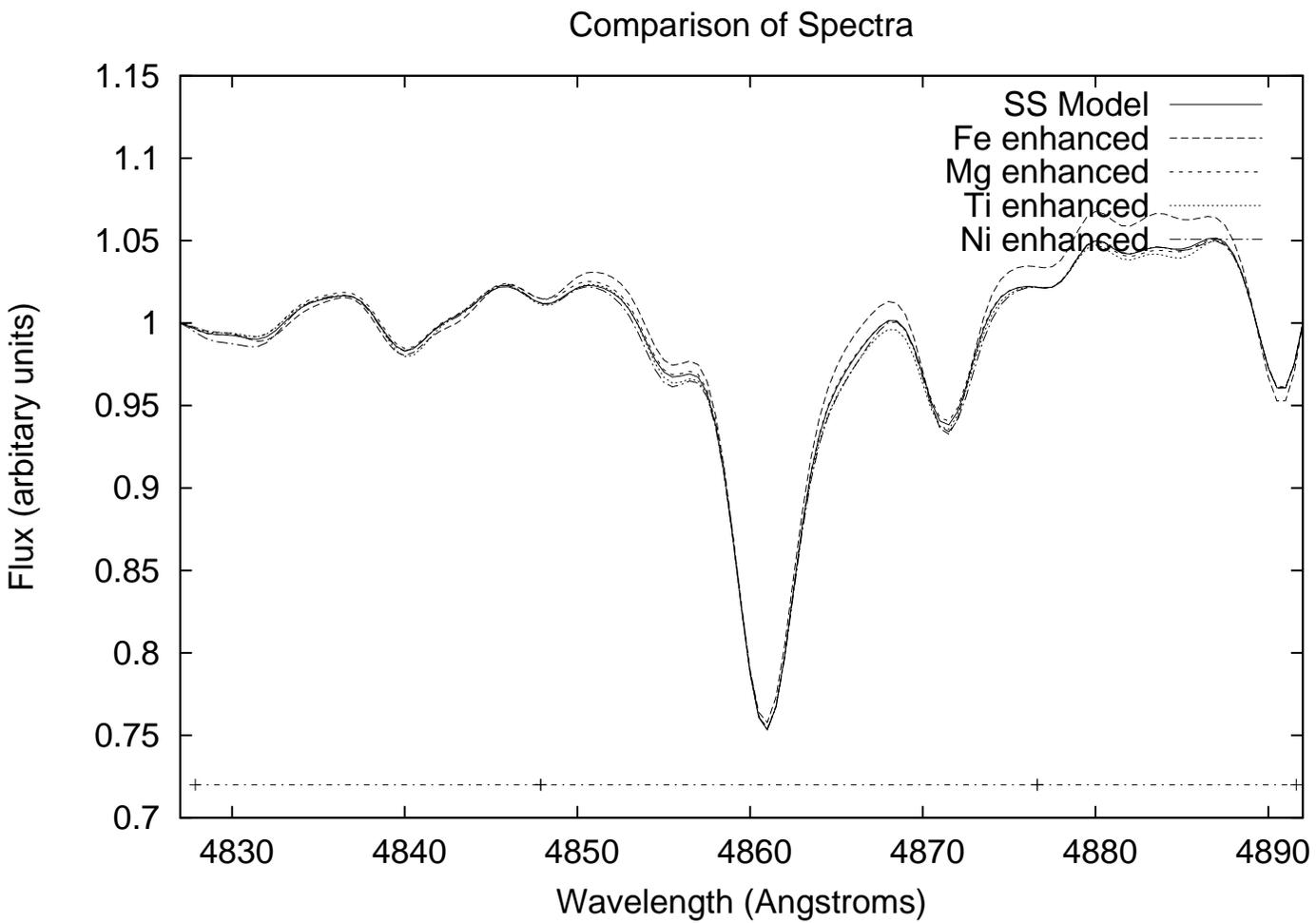}
\caption{Comparison of model spectra near the H-Beta line, with various
  element enhancements.  SS refers to scaled-solar, and the rest are enhanced
  by 0.3 dex, element by element, with total heavy element abundance held
  constant.  \label{4}}
\end{figure}

Figure \ref{3} should be compared to Figure \ref{4}, which shows several model
population spectra with [X/R]=0.3 dex, where X stands for Fe, Mg, Ti,
or Ni, and R stands for ``generic heavy element''. Other simple element
variations from solar were explored, but these four had the largest
impact on the spectrum shape.  We note that none of the elements effect
the model depth of the H$\beta$ line itself; only modest affects in
the continuum regions.  Visually, Fe enhancement raises the average
height of the red continuum band. Quantitatively, however, this ``extra
slope'' does little to change the actual index value. Furthermore, in
comparing model spectra with the observed globular cluster spectra,
raising the red continuum flux does not improve the appearance of the
spectral match.

A more quantitative way to analyze the effect due to element enhancement is by
looking at the spectral response of the indices when the various elements are
enhanced in the same way as \citet{s05}. The results of these calculations on
our model spectra can be found in Table 1. Row 1 is the model index while row
2 is the uncertainty assuming a $S/N = 100 $ at $5000$ \AA. Rows 3 through 25
list the change of index when the labeled element is enhanced by $0.3$ dex,
while the last row has all elements up by $0.3$ dex. As one can see, most
elements have little effect on the H$\beta$ index. Two iron-peak elements, Fe
and Ni, oppose each other in the sign of their effects, and two alpha
elements, Mg and Ti, oppose each other in the sign of their effects. If the
alpha elements and the iron peak elements internally rise or fall together,
then H$\beta$ is basically completely clean from spectral effects from element
enhancement. This is in broad agreement with observations made directly from
the model spectra.

Parenthetically, Table \ref{tab1} does not show similar cleanliness for any
other index, with Mg $b$ responding to Mg, Fe indices responding to Fe,
H$\delta$ responding to Fe and Si, and H$\gamma$ responding to C, O, Mg, Si,
Cr, and Fe!

The lack of signal in the H$\beta$ index seems to indicate that the reason for
larger spread in ages for grid with the H$\beta$ index as one of the axes is
most likely not due to enhancement of one element or any group of elements
that directly change the spectral shapes. There remains, perhaps, a
possibility of elements that do not make direct spectral changes, but might
affect the temperatures of the stars as a whole, such as O and the noble gases
He and Ne. An excess of O or a dearth of He would make the isochrones, at
least, around the main sequence turnoff region of the H-R diagram, cooler. If
metal rich globular clusters tend to have such a mixture, but elliptical
galaxies do not, then it may work out as observed, but of course there is no
reason to suspect a chemical bifurcation in the two classes of metal-rich
stellar populations.

\subsection{Horizontal Branch Morphology}

Horizontal branch morphology, that is ``red clump,'' ``extended,''
``blue,'' or ``extreme,'' is easy to determine with a good color
magnitude diagram of the stars within the globular cluster, but
difficult to disentangle via integrated light measures because of
significant degeneracy with both age and metallicity (cf. Fig 37 of
\citet{w94}). Blue Horizontal branch morphology can increase the
H$\beta$ index by as much as 0.75 \AA\ compared to clump
\citep{le00}. Fig 1 (a) illustrates how different horizontal
branch morphology can shift the model grids by showing two model grids with
different HB morphologies. This shift gives the appearance
that the globular clusters with blue horizontal branch morphologies are
younger or more metal poor than they really are. 

\begin{figure}
\epsscale{.80}
\plotone{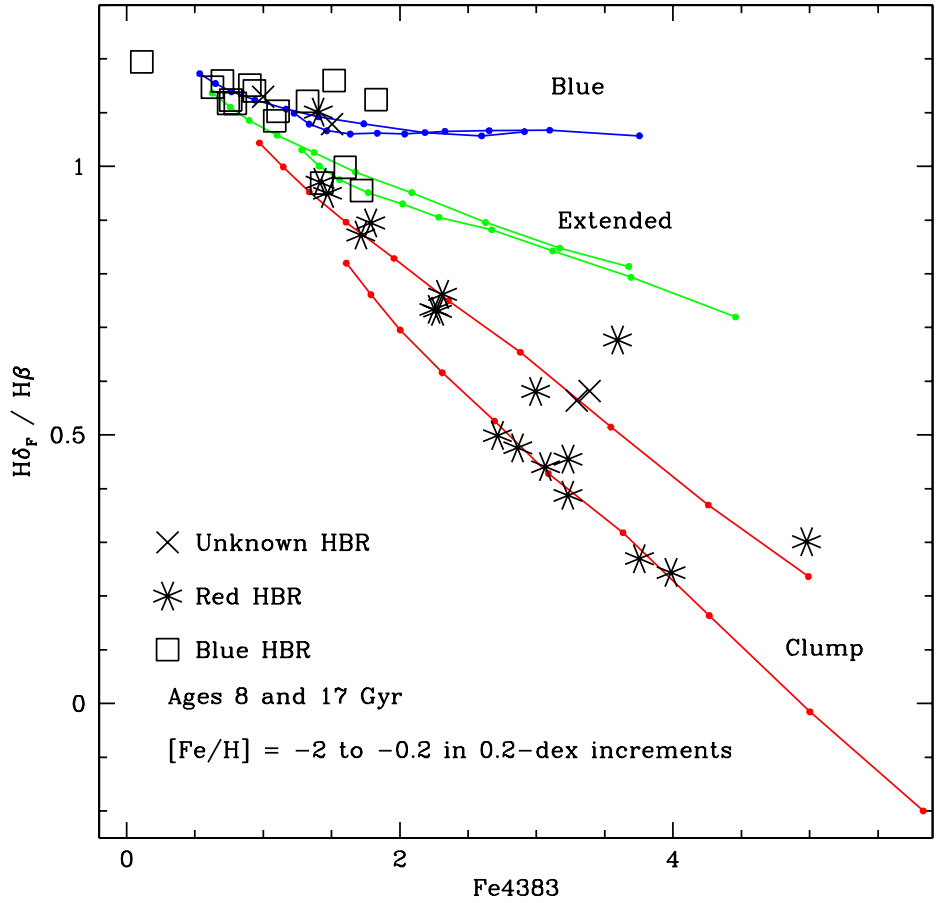}
\caption{Balmer index ratio vs. Fe4383, a horizontal branch diagnostic
  diagram. \citet{w94} Models for ages 8 and 17 Gyrs between [Fe/H] $-$2 and
  $-$0.2 are ploted in increments of 0.2 dex. The upper sequence has a horizontal
  branch morphology that is forced to be blue. The middle sequence represents
  a horizontal branch extended in temperature. The lower sequences represents
  a red clump mophology. Symbols for globular cluster data are as in previous
  figures.  \label{5}}
\end{figure}

\citet{sc04} proposed that the ratio of H$\delta_f$/H$\beta$ is more
sensitive to horizontal branch morphology than to age allowing us to
break the degeneracy present between these two parameters.  Graphs
with H$\delta_f$/H$\beta$ versus iron indices have globular clusters
with mostly blue horizontal branch morphologies that appear
displaced relative to the locus occupied by the models, as shown in
Fig. \ref{5}. Since in the Milky Way globular cluster system, the horizontal
branch morphology changes from blue to red at around [Fe/H] = $-1$, there is
some ambiguity with metallicity. 

It is unlikely that a ratio of Balmer indices
will be completely optimal for detecting an extended horizontal
branch. Indeed, we know of no effort to optimize the integrated-light
detection of horizontal branch types in the literature, so we provide
a simple one in this work. The idea is to pick three indices, and
force them to provide solutions for age, horizontal branch morphology,
and metallicity; three equations in three unknowns. The models used as
a base are the \citet{w94} models because the horizontal branch can be
forced to be clump, extended, or blue at any metallicity. Then an
error assigned to each color, magnitude, or index yields a propagated
error in age, morphology, or abundance.

The corners used for the linearization are ages 11 and 17 Gyr, [Fe/H]
of $-1.8$ and $-0.6$, and morphologies either red clump or blue, where clump
was assigned a numerical value of zero, and blue a numerical value of
one. The models also come in an ``extended'' horizontal
branch morphology that stretches from the red clump to log $T_{eff} = 4.0$,
but in this paper we only use those models for plots. The ``blue'' morphology
has stars between log $T_{eff} = 3.9$ and log $T_{eff} = 4.1$. Errors were
assigned to each index, and then used to do error propagation in the linear
solutions, giving a goodness-of-fit for age, morphology, and metallicity.

\begin{figure}
\epsscale{.80}
\plotone{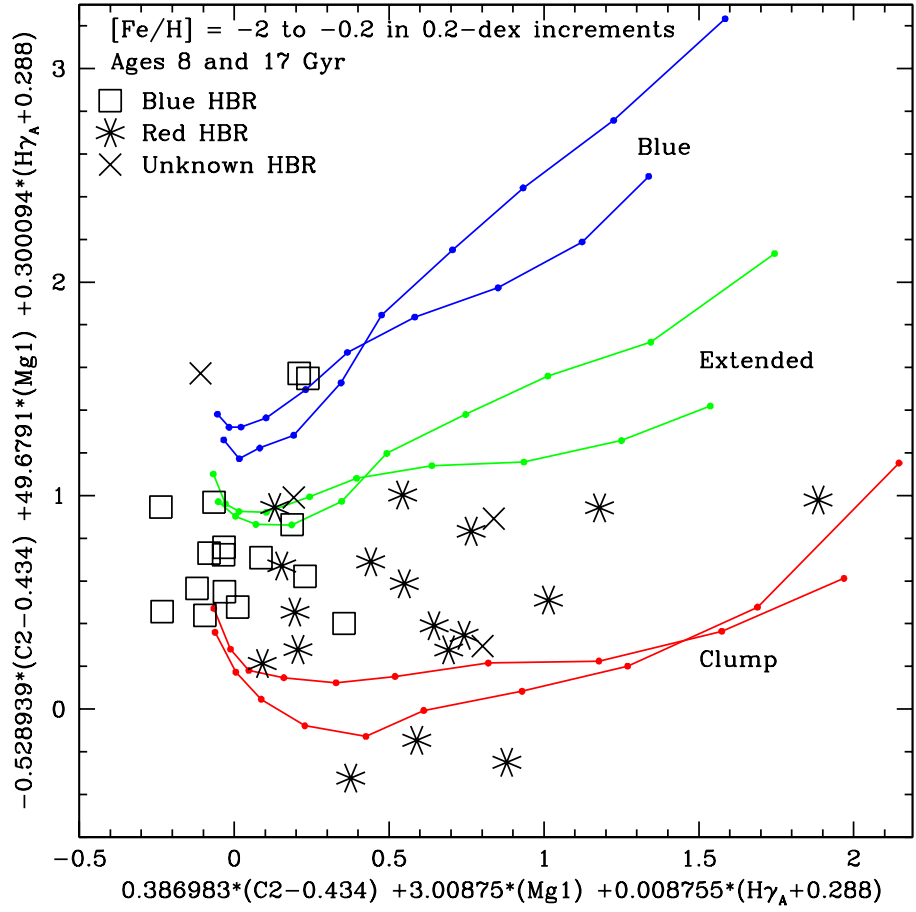}
\caption{Example diagnostic index combination plot that attempts to separate
  horizontal branch morphology (y-axis) and metallicity (x-axis). Model
  sequences are as in Figure \ref{5} \label{6}}
\end{figure}

An example of this is shown in Fig \ref{6}. For each trio of
indices, the horizontal branch solution is on the y axis, and the metallicity
solution is on the x axis. Model sequences for ages 8 and 17 Gyr are shown at
0.2-dex intervals from [Fe/H] = $-$2.0 to $-$0.2, and the collection of
\citet{sc04} indices are shown, symbols varying between HBR $< 0$ and HBR $>
0$, as in previous figures. In the figure the model sequences separate much
more than the simpler Balmer index ratio plot. However, we caution that the
models used are rather out of date, and better solutions might be found from
more up to date isochrones. 

In terms of horizontal branch morphology being responsible for the too-low
H$\beta$ strengths, this hypothesis seems doomed. The models that do not match
already have the reddest possible horizontal branch morphology, and the models
fit both blue and red horizontal branches for most clusters, just not the
anomalous group of redder clusters.

\subsection{Blue Stragglers}

No provision for blue straggler stars are in the synthesis models. If such
stars were present in the models, the Balmer line strengths would strengthen
by several tenths of \AA ngstroms, in a sense to make the low-lying globular
clusters lie even lower.

\subsection{Emission Fill in}

Filling in of the absorption lines due emission could change the depth of the
lines present in the spectra. This change in depth will lower the Lick/IDS
index value of the object which is what we want. Emission could be caused by
many things, such as: Gas clouds in the line of sight, planetary nebula,
supernova remnants, M dwarfs with active chromospheres, and AGB type stars.

For example, if we  adopt a star formation region-like recombination spectrum
for an optically thick nebula of $10^4$ K and  $10^4$ electrons per cubic
centimeter, \citet{o89} gives the relative Balmer line intensities of
$j{\gamma}/j{\beta} = 0.469$ and $j{\delta}/j{\beta} = 0.260$. For a given
H$\beta$ index fill-in value, the equivalent widths of the H$\gamma$ and
H$\delta$ indices can be predicted after correcting for (1) underlying
continuum shape and (2) the widths of the indices themselves. Using values of
$F_{c,\ \gamma}/F_{c,\ \beta} = 0.84$ and $F_{c,\ \delta}/F_{c,\ \beta} = 0.81$
for the relative continuum flux ratios, a one-Angstrom fill-in of H$\beta$
propagates to fill-ins for the higher-order indices of $\Delta {\rm H}\gamma_F
= 0.76$ \AA , $\Delta {\rm H}\gamma_A
= 0.37$ \AA , $\Delta {\rm H}\delta_F
= 0.44$ \AA , and $\Delta {\rm H}\delta_A
= 0.24$ \AA .

Planetary nebulae can be seen, one by one, in Milky Way globular clusters, and
only M5 has a planetary nebula, so they should be rare in M31 globular
clusters as well. Supernova remnants are much more improbable. Gas clouds
containing neutral sodium are known to exist along most lines of sight out of
the galaxy \citep{aud} but there is no reason to expect ionized hydrogen to
linger in the potential wells of globular clusters since the RMS velocity for
a 10,000 K proton is $v_{\rm rms} = ( 3kT/m )^{1/2} \sim 15$  km s$^{-1}$
exceeds the escape velocity of all but the largest globular clusters. We thus
discount these three explanations in general, keeping in mind that specific
globular clusters can be affected this way. 

However, the stellar sources are less easy to discount. We discuss active cool
dwarfs and flaring giants together. M dwarfs are known to have active
chromospheres, although old ones get less active \citep{west}. In addition,
\citet{sc05} observed a probable bright, red giant flaring in their
spectra. This star is bright enough so that, by itself, it will alter the
integrated Balmer line strengths. The character of these two sources is
different, however, in that the numerous M dwarfs are spatially broad and
should be nearly constant in Balmer emission output while the giants are
spatially discrete, and should be highly time-variable. M dwarf light will
still give a net Balmer emission signal because the M dwarfs are concentrated
toward the center of the cluster, albeit less so than the more massive
stars. In support of cool giants causing fill-in, the most metal-poor clusters
do not have very cool giants. It is only at [Fe/H] $\approx -1$ and above that
clusters begin to have long period variables and genuine M-type giants, and
these are the clusters that show the anomalously low H$\beta$ strengths.

Under some extreme assumptions, we use our models and the data of
\citet{kafka06} to estimate the contribution of the emission of active M
dwarfs. The \citet{cohen98} definition of H$\alpha$ is output by our
models. The \citet{kafka06} definition is somewhat different, but of course
quite similar. For cooler M dwarfs, the index itself goes negative
(emission-like) due to TiO absorption, and we were able to confidently trace
the non-active envelope in the \citet{kafka06} data, and then assign an
apparent turn-on temperature of 3600 K, and a ballpark ``fully active''
$\Delta$H$\alpha \approx -4.0$ \AA\ of equivalent width. Assuming that 100\%
of the stars cooler than 3600 K were fully active, we recalculated the
models. We also assumed that the initial mass function was a power-law all the
way to a cutoff of $0.1 M_\odot$, which makes the coolest dwarfs more
important in integrated light than direct counting suggests. The H$\alpha$ and
inferred H$\beta$ results are listed in Table \ref{tab2} as a function of
metallicity. 

The $\Delta$H$\beta$ values in Table \ref{tab2}, even inflated as they are,
are still short of the approximately 0.5 \AA\ needed to come close to solving
our H$\beta$ dilemma, but the generic behavior is interesting, namely that
metal-poor populations have so few stars that cool that the emission is
completely negligible, but then the metal-rich populations, as judged by the
last entry in the table, seem to saturate or plateau (because more and more
non-active but cool stars contribute to the flux). This leads to a sort of on
or off state, with ``on'' happening for old populations more metal rich than
about $-0.5$ dex in [Z/H].

This tends to lend support to the cool-giant hypothesis, since they would
share a similar gross temperature behavior with metallicity as do the dwarfs,
and the metal-poor clusters tend to lie on the old-age model grid. This does
imply some things for elliptical galaxies, however, if it were true. The
galaxies would have such large numbers of stars in a spectroscopic aperture
that stochastic fluctuations in AGB star activity would be minimal, so they
would reach an average H$\alpha$ and H$\beta$ value with minimal scatter. The
metal-rich portions of their stellar populations would contribute a
partially-infilled Balmer index series, and so the larger the percentage of
their population that is metal-rich, the younger they would appear in a
Balmer-metal diagram. While something like this trend is seen observationally
\citep{wor95} one should not jump to conclusions since even the higher-order
Balmer features show a similar behavior.

\subsection{Models}

Are the line depths of current sets of models too deep? The Worthey models
agree quite well with more modern model sets, especially after rescaling the
ages by subtracting 2 Gyr. If, however, all authors are making the same
mistake and all Balmer line strengths should be dropped to a level to make the
low-lying globular clusters fit along the old-age sequence, then there are
some implications. First, even with correction for horizontal branch
morphology, the metal-poor clusters will still look substantially
younger. Second, the average elliptical galaxy will look young enough to raise
eyebrows.

Despite this quandary, there may be one unlikely way to get the models to fit
everything, or nearly everything, and that is to invoke hefty abundance ratio
systematics, especially with oxygen and helium, that we could not test
effectively in this paper. Such a scheme would require that element ratios
drift in opposite directions for metal-rich globular clusters versus
elliptical galaxies. The effects of O and He abundance would have operate
mostly on the isochrone temperatures and age scales, and not operate
significantly on the integrated stellar spectra. However, in the absence of
more direct observational evidence, this scheme is very speculative.

\section{Conclusion}

Extracting information from the integrated light of stellar populations is not
an easy process, since there are many complex factors affecting the
spectra. Decoupling the age-metallicity degeneracy by graphing Balmer line
indices vs. metal feature indices has allowed us to learn much more about
stellar populations, however the weakness of the observed H$\beta$ line
relative to the models needs to be explained. This paper shows that altered
abundance ratios, are unable to account for the observed weakness in the
Balmer line strengths of globular clusters.
 
Of the other factors that can potentially affect the Balmer line strengths,
horizontal branch morphology effects are hard to disentangle since they are
also degenerate with age and metallicity, but seem well understood. By marking
the clusters that have extended or blue horizontal branches it becomes clear
that horizontal branch morphology cannot solve the H$\beta$ problem.
Likewise, inclusion of blue stragglers will not help, even if there was
evidence for a strong modulation of blue straggler frequency with metallicity,
which there is not \citep{sand05}. Emission fill-in of the Balmer lines due to
hydrogen recombination lines from external nebulae is probably ruled out,
except for case like the planetary nebula in M5.  Stellar activity in
individual cluster stars seems to be the only surviving mechanism that has
good evidence. However, even being generous, the cool tail of M dwarfs do not
appear to be able to generate enough flux to cause the modulation in H$\beta$
needed. The remaining stellar source is flaring in M giants. These stars are
bright enough, and inherently stochastic in nature, which seems to fit the
observations of clusters that scatter to low H$\beta$ rather randomly.
Finally, it remains a long-shot possibility that abundance ratios in O or the
noble gases can cause isochrone temperature drifts severe enough to affect the
H$\beta$ problem.

\acknowledgments

The authors gratefully acknowledge support from National Science
Foundation grants AST-0307487 and AST-0346347.

\clearpage

\clearpage

\clearpage

\clearpage

\begin{table}
\caption{Spectral response of indices under various element
  enhancements \label{tab1}} 
\begin{tabular} {| l | r | r | r | r | r | r |}
\hline
      & H$\beta$ & H$\delta_F$ & H$\gamma_F$ & Mg$_b$ & Fe5270 & Fe5335 \\
\hline
$I_{OB}$ & 1.522 & 0.613 & -0.709 & 2.121 & 2.025 & 1.762 \\
error & 0.138 & 0.119 & 0.121 & 0.155 & 0.173 & 0.199 \\
C & 0.00 & -0.16 & -4.17 & -0.19 & 0.16 & 0.06 \\
N & 0.00 & -0.04 & -0.03 & -0.01 & 0.06 & 0.01 \\
O & 0.04 & -0.11 & 0.76 & 0.12 & -0.04 & -0.01 \\
Na & 0.01 & -0.01 & 0.07 & -0.09 & -0.03 & -0.03 \\
Mg & -0.29 & 0.16 & 1.12 & 4.83 & -0.32 & -0.26 \\
Al & 0.02 & 0.01 & 0.11 & -0.06 & -0.05 & -0.05 \\ 
Si & 0.07 & 1.88 & 0.79 & -0.32 & -0.09 & -0.07 \\
S & 0.00 & 0.00 & 0.01 & 0.00 & 0.00 & 0.00 \\
K & 0.00 & 0.00 & 0.01 & 0.00 & -0.01 & -0.01 \\
Ca & -0.02 & 0.56 & -0.21 & 0.06 & 0.06 & 0.03 \\
Sc & -0.01 & -0.03 & -0.27 & 0.00 & -0.14 & 0.03 \\ 
Ti & 0.28 & -0.54 & -0.10 & 0.01 & 0.28 & 0.14 \\
V & -0.02 & 0.53 & -0.02 & -0.02 & -0.05 & 0.01 \\
Cr & -0.12 & 0.03 & 0.66 & -0.86 & 0.10 & 0.39 \\
Mn & -0.02 & -0.41 & -0.04 & -0.09 & 0.09 & 0.04 \\
Fe & -0.57 & -2.62 & -0.85 & -0.79 & 1.88 & 1.54 \\
Co & -0.02 & -0.21 & -0.01 & 0.00 & 0.13 & 0.16 \\
Ni & 0.61 & -0.09 & -0.11 & 0.00 & 0.06 & 0.00 \\
Cu & 0.00 & 0.00 & 0.00 & -0.07 & -0.01 & 0.00 \\
Zn & 0.00 & 0.00 & 0.00 & 0.00 & 0.00 & 0.00 \\
Sr & 0.00 & -0.34 & 0.00 & 0.00 & 0.00 & 0.00\\
Ba & 0.00 &-0.06 & 0.00 & 0.00 & 0.00 & 0.00 \\
Eu & 0.00 & -0.06 & 0.00 & 0.00 & 0.00 & 0.00 \\
upX2 & 0.09 & 1.95 & 2.31 & 4.48 & 0.45 & -0.21 \\
\hline
\end{tabular}

\tablecomments{Row 1 is the model index, row 2 is the uncertainty assuming a
  $S/N = 100 $ at $5000$ \AA, rows 3 through 25 list the change of index when
  the labeled element is  enhanced by $0.3$ dex, and the last row has all
  elements up by $0.3$ dex.} 
\end{table}

\begin{table}
\caption{$\Delta$ Balmer indices for active M dwarf experiment as a function of
  [Fe/H] \label{tab2} } 
\begin{tabular} {lrr}
\hline
[Fe/H] & $\Delta$H$\alpha$ & $\Delta$H$\beta$ \\
\hline
$-2.0$ & 0.013 & 0.004 \\
$-1.5$ & 0.021 & 0.006 \\
$-1.0$ & 0.033 & 0.010 \\
$-0.5$ & 0.155 & 0.056 \\
$\ 0.0$ & 0.307 & 0.126 \\
$\ 0.5$ & 0.278 & 0.131 \\

\hline
\end{tabular}

\tablecomments{The model sequence is for a 10 Gyr age simple stellar
  population. The IMF was set to a power law with a lower mass cutoff of 0.1
  $M_\odot$. This, plus the assumption that every dwarf with $\Theta > 0.42$
  is highly active, both will tend to exaggerate the effects of including
  active M dwarfs. The H$\alpha$ definition used is that of \citet{cohen98}
  and the H$\beta$ definition is that of \citet{w94b}. }
\end{table}

\clearpage

\end{document}